\documentclass[a4paper,fleqn,usenatbib]{mnras}
\hypersetup{draft}
\pdfoutput=1
\usepackage{newtxtext,newtxmath}
\usepackage[T1]{fontenc}
\usepackage{ae,aecompl}
\usepackage{graphicx}
\usepackage{amsmath}
\usepackage{amssymb}
\usepackage{tablefootnote}

\newcommand{\af}{[$\alpha$/Fe] }
\newcommand{\afe}{[$\alpha$/Fe]}

\newcommand{\vel}{km s$^{-1}$ }

\newcommand{\gai}{\textit{Gaia} }

\title[Chemodynamics of the Solar neighbourhood]{The GALAH Survey: Chemodynamics of the Solar Neighbourhood}

\author[M. R. Hayden et al.]{
Michael R. Hayden$^{1,2}$\thanks{E-mail: michael.hayden@sydney.edu.au)},
Joss Bland-Hawthorn$^{1,2}$,
Sanjib Sharma$^{1,2}$,
Ken Freeman$^{3}$,\newauthor
Janez Kos$^{1,4}$,
Sven Buder$^{5}$,
Borja Anguiano$^{6}$,
Martin Asplund$^{3,2}$,
Boquan Chen$^{1,2}$,\newauthor
Shourya Khanna$^{1}$,
Jane Lin$^{3}$,
Jonathan Horner$^{7}$,
Sarah Martell$^{8,2}$,
Rosemary Wyse$^{9}$,\newauthor
Daniel Zucker$^{10}$,
Tomaz Zwitter$^{4}$
\\
$^{1}$Sydney Institute for Astronomy, School of Physics, University of Sydney, NSW 2006, Australia\\
$^{2}$ARC Centre of Excellence for All Sky Astrophysics in 3D (ASTRO-3D), Australia\\
$^{3}$Research School of Astronomy \& Astrophysics, Australian National University, ACT 2611, Australia\\
$^{4}$Faculty of mathematics and physics, University of Ljubljana, Jadranska 19, 1000 Ljubljana, Slovenia\\
$^{5}$ Max Planck Institute for Astronomy (MPIA), Koenigstuhl 17, 69117 Heidelberg, Germany\\
$^{6}$ Department of Astronomy, University of Virginia, Charlottesville, VA 22904-4325, USA \\
$^{7}$ Centre for Astrophysics, University of Southern Queensland, Toowoomba, Qld 4350, Australia\\
$^{8}$ School of Physics, University of New South Wales, Sydney, NSW 2052, Australia\\
$^{9}$ Johns Hopkins University, Dept of Physics \& Astronomy, Baltimore, MD 21218 \\
$^{10}$ Department of Physics and Astronomy, Macquarie University, Sydney, NSW 2109, Australia\\
}

\date{}
\pubyear{2019}
\begin{document}
\label{firstpage}
\pagerange{\pageref{firstpage}--\pageref{lastpage}}
\maketitle

\begin{abstract}
We present the chemodynamic structure of the solar neighbourhood using 62 814 stars within a 500 pc sphere of the Sun observed by GALAH and with astrometric parameters from \gai DR2. We measure the velocity dispersion for all three components (vertical, radial, and tangential) and find that it varies smoothly with [Fe/H] and \af for each component. The vertical component is especially clean, with $\sigma_{v_z}$ increasing from a low of $8$ \vel at solar-\af and [Fe/H] to a high of more than $50$ \vel for more metal-poor and \af enhanced populations. We find no evidence of a dramatic decrease in the velocity dispersion of the highest-\af populations as claimed in surveys prior to \gai DR2, although the trend of increasing velocity dispersion with \af for the same metallicity does significantly flatten at high-\afe. The eccentricity distribution for local stars varies most strongly as a function of \afe, where stars with \af$<0.1$ dex having generally circular orbits ($e<0.15$), while the median eccentricity increases rapidly for more \af enhanced stellar populations up to $e\sim0.35$. These \af enhanced populations have guiding radii consistent with origins in the inner Galaxy. Of the stars with metallicities much higher than the local ISM ([Fe/H]>0.1 dex), we find that more than 70\% have $e<0.2$ and are likely observed in the solar neighbourhood through churning/migration rather than blurring effects, as the epicyclic motion for these stars is not large enough to reach the radii at which they were likely born based on their metallicity.
\end{abstract}




\section{Introduction}

Galactic Archaeology is the study of the structure and history of the Milky Way with the aim of understanding galaxy evolution using the fully resolved stellar inventory, from the long-lived, low mass dwarfs to the most massive, hot young stars. The Milky Way provides a unique test bed for many theories of galaxy evolution due to the ability to resolve individual stars and stellar populations, which is generally not possible in external galaxies. Stars contain the chemical fingerprint of the gas from which they formed, and can be used as fossil records to look back in time and see how the ISM of the Galaxy has evolved. Despite this, there is much debate about the different structural properties of the Galaxy as a whole (e.g., \citealt{rix2013,jbh2016}). The primary cause of that debate are the poor observational constraints on the various models, which is a problem even for the Solar neighbourhood.

The chemodynamic structure of the solar neighbourhood is a complex mesh of many different stellar populations that have been built up over the history of the Milky Way. The stars that are found close to the Sun today span a huge range in age and come from a variety of birth radii. This is most easily demonstrated by the large number of super-solar metallicity ([Fe/H]$>0.1$ dex) stars observed in the solar neighbourhood and lack of an overall age-metallicity relationship (e.g., \citealt{casagrande2011,haywood2013,bergemann2014,kordopatis2015,buder2018s,hayden2018}). Stars with [Fe/H]$>0.1$ dex are too metal-rich to have formed in the solar neighbourhood as the local ISM metallicity today is well established from the B star population at slightly subsolar values \citep{nieva2012}. The likely origin of these stars is the inner Galaxy, as there are strong radial metallicity gradients in the disk where the median metallicity of stars in the inner Galaxy is much higher than in the solar neighbourhood (e.g., \citealt{luck2011b,lemasle2013,anders2014,hayden2014}).

The local ISM appears to be fairly well mixed, with young population tracers such as B stars and Cepheids showing only small scale variation over a large range in azimuth (\citealt{luck2011a,nieva2012}, but see also \citealt{balser2011,balser2015} that do find variations in HII regions with azimuth). How super-solar metallicity stars reach the solar neighbourhood is an open question. Radial mixing processes have been proposed to explain the chemodynamic properties of the Galaxy (e.g., \citealt{sellwood2002,schonrich2009,minchev2010}), which are differentiated by the effect on the angular momentum of the orbits. In blurring, stellar orbits become heated with time and have large epicyclic motions over their orbit, but the angular momentum remains conserved. In churning (migration), there is a change in angular momentum of an orbit due to interactions associated with non-axisymmetric structure (i.e., spiral arms or the bar). Changing the angular momentum of an orbit causes a change in guiding radius, and if the interaction is due to a corotation resonance, other parameters such as eccentricity are preserved. Stars are equally likely to migrate in or out. If spiral arms are transient, the net effect is to move stars to larger radii due to the exponential density law of the stellar disk.  

The velocity dispersions of stellar populations in the solar neighbourhood contain potential hints of a major merger early on in the history of the Galaxy (\citealt{minchev2014}). The velocity dispersion is generally observed to increase with age, as older populations get heated with time or were potentially born in a kinematically hotter environment than the present star forming disk, which has a scale height of $<100$ pc \citep{jbh2016}. However, observations using data from RAVE have shown a downturn in the velocity dispersion for the highest-\af (i.e., likely the oldest) stellar populations across a range of metallicities in the solar neighbourhood \citep{minchev2014}. This same signature was observed in the simulations of \citet{minchev2013}, who found that a major merger at $z\sim2$ drove enhanced radial migration of stellar populations to the solar neighbourhood. Migration can cool the disk as stars that migrate often have cooler kinematics than stars of the same age that are born locally (e.g., \citealt{minchev2012,vera-ciro2014}). This result was less obvious in observations from the Gaia-ESO survey \citep{guiglion2015,hayden2018}, which showed a shallower turnover in the velocity dispersion for higher-\af populations compared to the RAVE results.

Our present understanding of the Milky Way is in the midst of a revolution. With the advent of large-scale spectroscopic surveys such as SEGUE \citep{yanny2009}, RAVE \citep{steinmetz2006}, LAMOST \citep{deng2012}, APOGEE \citep{majewski2017}, GALAH \citep{desilva2015}, and future surveys such as 4MOST \citep{dejong2014} and WEAVE \citep{dalton2014}, spectroscopic observations will be available for millions of stars in all Galactic components. Combined with the addition of precise astrometric information from the \gai satellite \citep{perryman2001}, we will have the ability to determine the precise kinematic, dynamic, and temporal structure of stellar populations throughout the Galaxy. \gai alone has found an incredible amount of substructure in velocity and action space in the solar neighbourhood (e.g., \citealt{antoja2018,trick2018}). The precise kinematic properties estimated by \gai used in conjunction with detailed chemistry estimates from the large spectroscopic surveys mentioned above provide the best picture to date of the structure and evolution of the Milky Way.

For this paper, we use observations of tens of thousands of stars found within 500 pc of the solar position from GALAH DR2 \citep{buder2018dr2} and astrometric information from \gai DR2 \citep{gaia2018dr2} to unravel the structure of the local disk. GALAH is a high resolution ($R\sim28 000$) spectroscopic survey of nearly one million stars throughout the Milky Way, which provides precise atmospheric parameters and abundances for dozens of elements. Used in conjunction with proper motions and parallax measurements from \gai DR2, we are able estimate orbital properties for the entire GALAH solar neighbourhood sample to map out the chemodynamic structure of the local disk. This paper is organised as follows: in Section 2 we discuss the data used in our analysis, in Section 3 we present our results on the velocity dispersion and orbital properties as a function of chemistry for stars near the Solar position, and in Section 4 we discuss our findings in the context of the chemodynamic structure of the solar neighbourhood, before drawing our conclusions in Section 5.
\vspace{3cm}
\section{Data}

\begin{table}
\centering
\caption{Data cuts used on the GALAH DR2 \gai DR2 sample.}
\label{datacuts}
\begin{tabular}{cc}
\hline
Flag & Value \\
\hline
cannon\_flag \tablefootnote{a description of the data flag bitmasks can be found here: https://datacentral.org.au/docs/pages/galah/table-schema/dr2-table-schema/} & 0 \\
abundance\_flag$^1$ & $<3$ \\
SNR & $>20$ \\
$\sigma_{\mathrm{T_{eff}}}$ & $<200$ K \\
$\mathrm{[Fe/H]}$ &  $-1.0 < \mathrm{[Fe/H]} < 0.5$ dex\\
d$_s$ & $<500$ pc \\
\hline
\end{tabular}
\end{table}

\begin{figure*}
\centering
\includegraphics[height=3.2in]{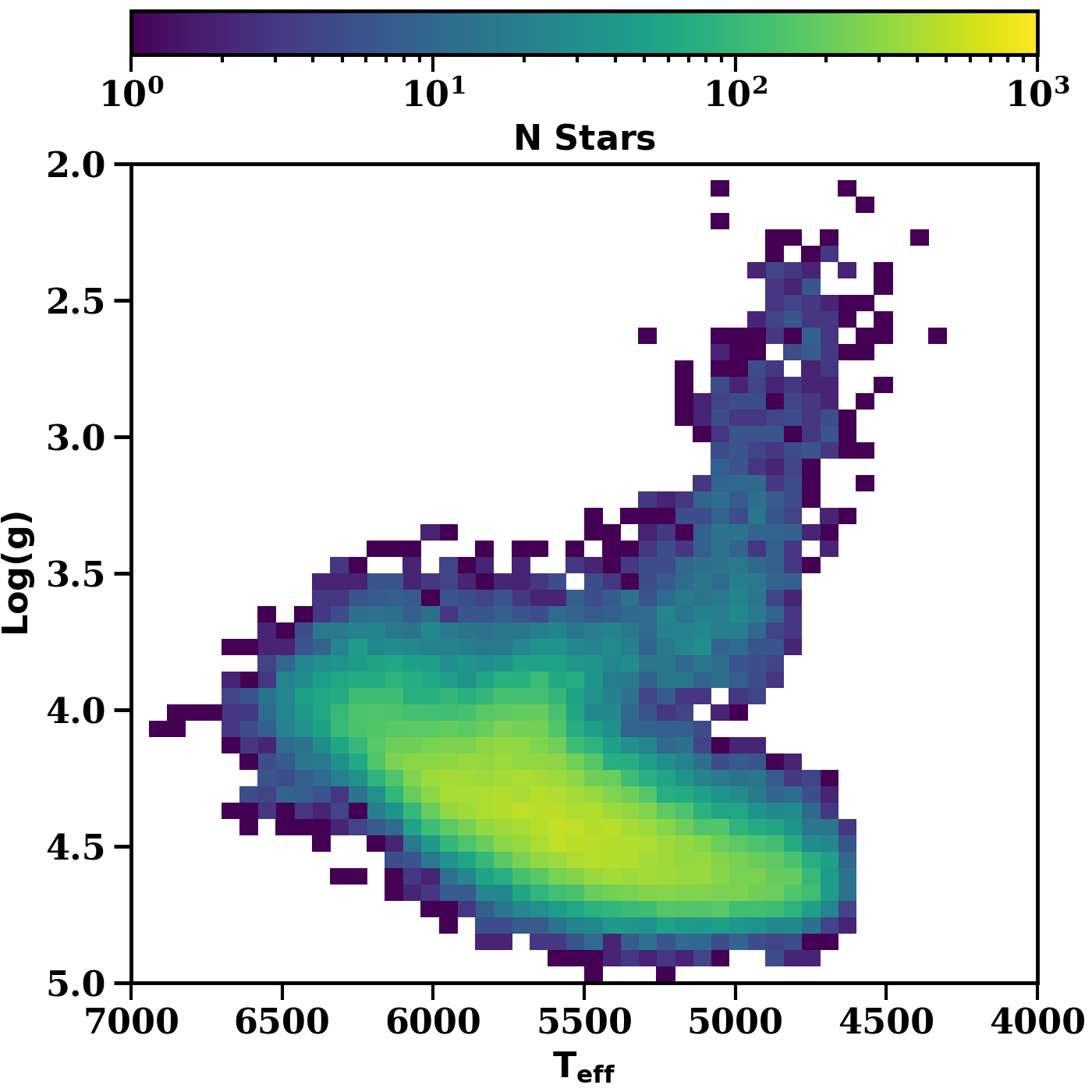}
\includegraphics[height=3.2in]{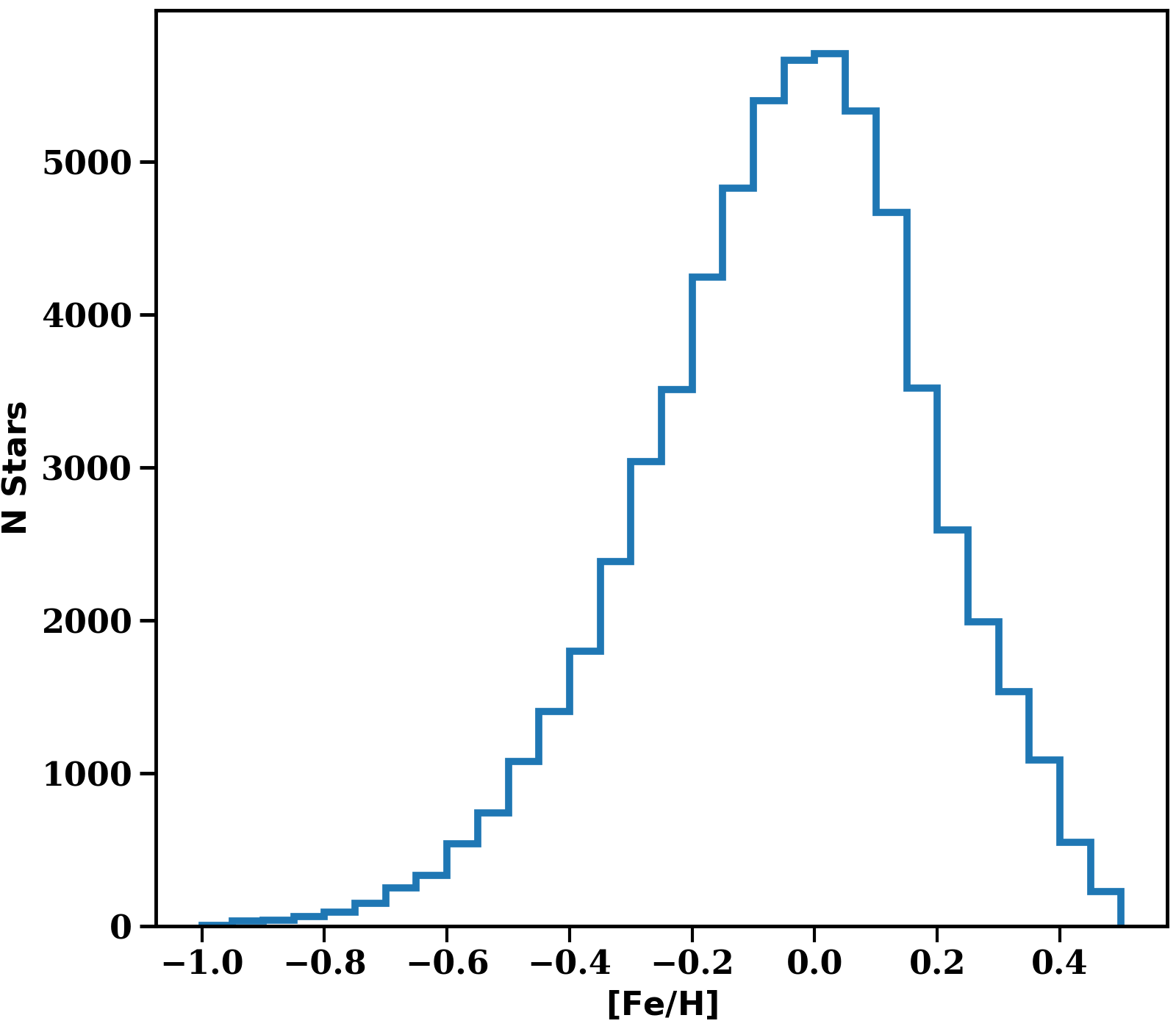}
\caption{\textbf{Left:} The H-R diagram for our sample of 62 814 GALAH stars within 500pc of the solar position. \textbf{Right:} The metallicity distribution function of the sample. Note the large fraction of super-solar metallicities stars despite the local ISM being $\sim$solar metallicity.}
\label{hrdiagram}
\end{figure*}

\begin{figure}
\centering
\includegraphics[width=3.5in]{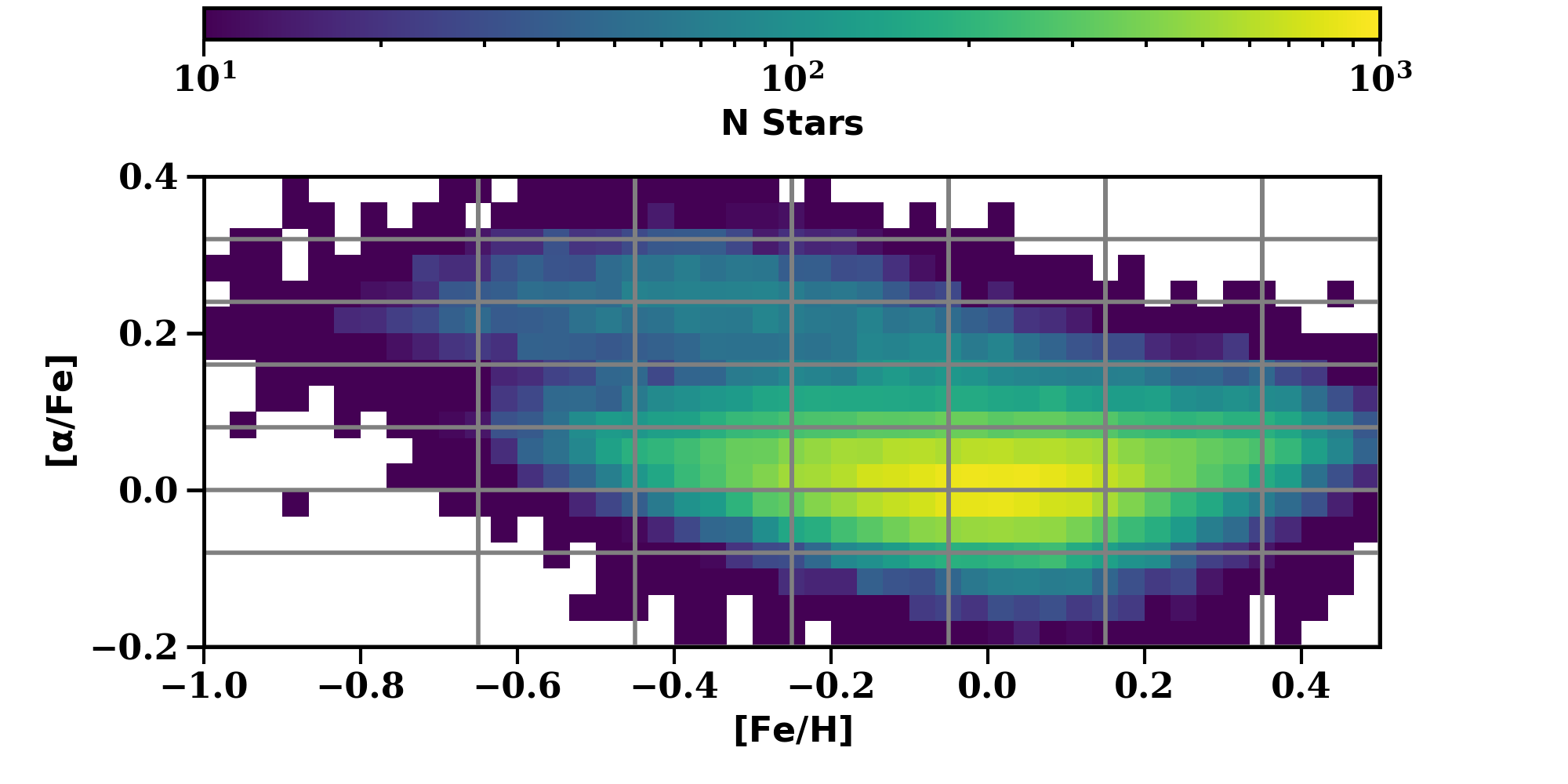}
\caption{The [$\alpha$/Fe] vs. [Fe/H] plane for the GALAH sample within 500 pc of the solar position. Grey lines denote the bins in metallicity and \af used later in this paper.}
\label{afsample}
\end{figure}

Spectroscopic data are taken from the internal GALAH DR2 release, which contains the public DR2 \citep{buder2018dr2}, and also additional fields for K2 \citep{wittenmyer2018} and TESS \citep{sharma2018} follow-up that contain large numbers of stars in the solar neighbourhood. GALAH uses the High Efficiency and Resolution Multi-Element Spectrograph (HERMES, \citealt{sheinis2015}) instrument, which is a high resolution ($R\sim28 000$) multi-fibre
spectrograph mounted on the 3.9 meter Anglo Australian Telescope (AAT). HERMES covers four wavelength ranges ($4 713-4 903${\AA}, $5 648-5 873${\AA}, $6 478-6 737${\AA}, and $7 585-7 887${\AA}), carefully selected to maximise the number of elemental abundances that are able to be measured. Observations are reduced through a standardised pipeline developed for the GALAH survey as described in \citet{Kos2017}. Stellar atmospheric parameters and individual abundances are derived using a combination of SME \citep{valenti1996,piskunov2017}, to develop a large training set covering parameter space, which is then fed into the Cannon \citep{ness2015} to obtain estimates for the entire sample. The precision of individual abundances [X/Fe] is typically $\sim0.05$ dex, while the random errors in radial velocities are $\sim100\ \mathrm{m\ s^{-1}}$ \citep{Zwitter2018}. For \af, we use an average of the Ti, Mg, and Al abundances from GALAH DR2. Data quality cuts are listed in Table \ref{datacuts}.
As we are interested primarily in the chemodynamics of the disk, we restrict our sample to $-1.0 < \mathrm{[Fe/H]} < 0.5$ dex in an attempt to minimise the impact of the halo in our analysis, in particular estimates of the velocity dispersion. 

Astrometric parameters are taken from \gai DR2 \citep{gaia2018dr2}, which have random errors on the order of a few \% for the local sample used in this paper. We restrict our sample to stars within 500 pc, with the aim of studying the stellar populations of the local volume. This sample contains 62 814 stars which span a range of stellar parameters and metallicities, as shown in Fig. \ref{hrdiagram}. The local volume is dominated by dwarf stars. An interesting feature of the metallicity distribution function (MDF), here measured using the iron abundance relative to solar values [Fe/H] \citep{asplund2009}, is that 45\% of stars in the sample are solar metallicity or above, and 25\% have metallicities $>0.1$ dex. The distribution of stars in the \af vs. [Fe/H] plane, showing the clear separation between the chemical thick and thin disks at lower metallicities, is shown in Fig. \ref{afsample}. The grey lines denote the different metallicity and \af bins used in our analysis. With the exception of the lowest \af bin, the sample is divided in \af at fixed increments of 0.08 dex and the zero-points chosen such that the overlap between the chemical thick and thin disks is minimised. The lowest-\af bin has a width of 0.12 dex. The metallicity bins are spaced by 0.2 dex, with the exception of the most metal-rich and metal-poor bins.

\subsection{Distances and Orbits}

\begin{figure}
\centering
\includegraphics[width=3.2in]{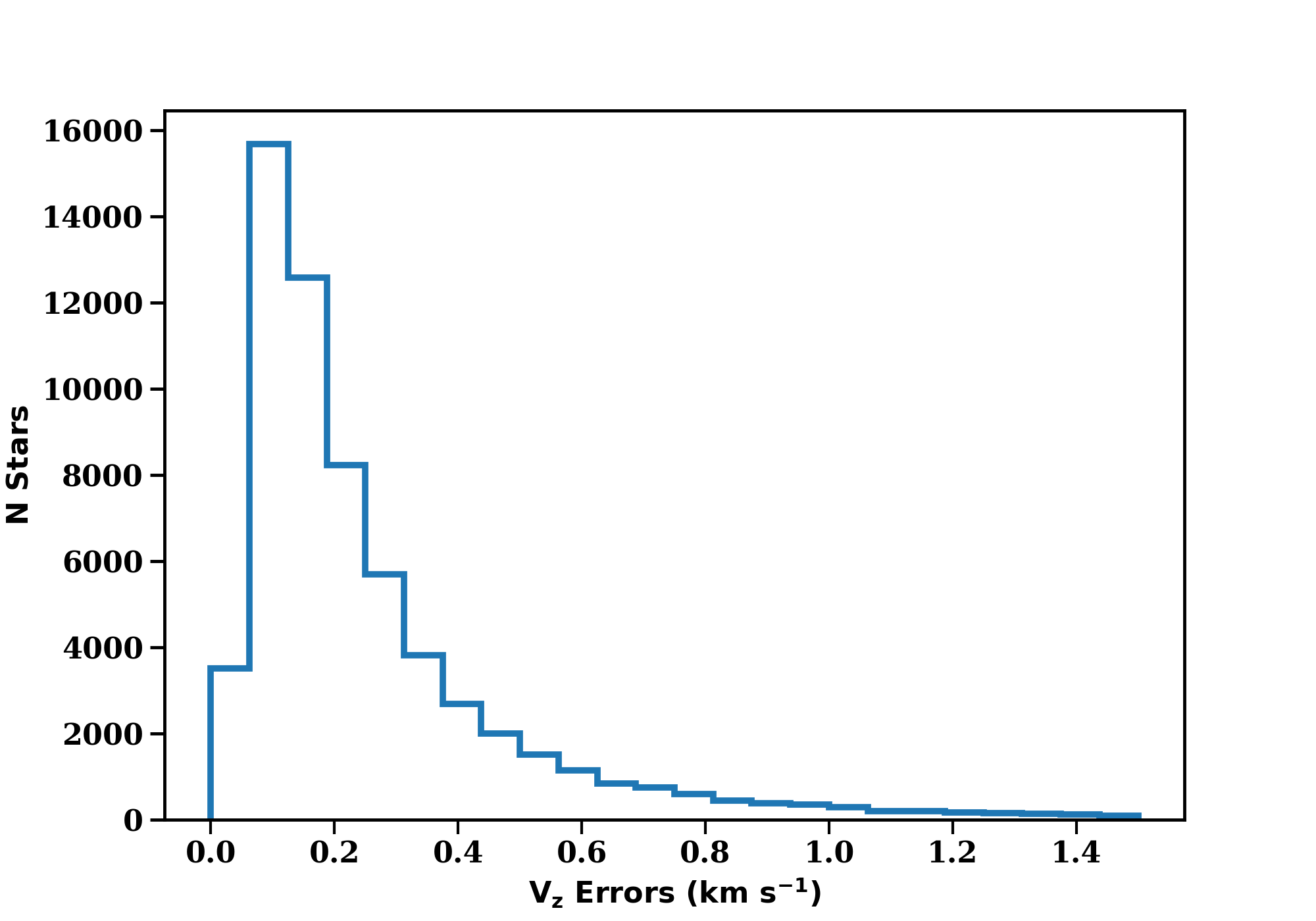}
\caption{The distribution of errors in the vertical velocity component. Errors in the radial and tangential components are comparable to those in the vertical component. Velocities are determined with proper motions from \gai DR2 and radial velocities from GALAH DR2, while errors are estimated via 1 000 Monte Carlo runs from the input parameters and their uncertainties. The errors in this sample for the individual velocity components are an order of magnitude improvement over previous studies. }
\label{vsample}
\end{figure}

The distances to stars are derived using the estimated parallax from \gai DR2 and a Bayesian technique similar to that outlined in \citet{bailer-jones2015}. We assume a simple prior with a single exponential disk that has a scale height of 300 pc and a scale length of 2.7 kpc, which is used to generate a probability distribution (PDF) in distance space. The distance is characterised by the mode of the PDF, while the errors in distance are estimated using the 16th and 84th percentiles of the PDF. As the sample is restricted to stars within 500 pc of the Sun, the parallax accuracy is high ($<10$\% error).

Kinematic and orbital properties for each star are derived with \textit{Galpy} \citep{bovy2015} using the St\"ackel analytic approximations outlined in \citet{mackereth2018} and the default ``MWHaloPotential2014'', rescaled such that $R_{\odot}=8.2$ kpc and the local circular velocity is 238 km s$^{-1}$ \citep{jbh2016}. 
The solar motion relative to the local standard of rest (LSR) of (11.65, 12.24, 7.25) (U$_\odot$,V$_\odot$,W$_\odot$) is used \citep{schonrich2010}. 1 000 MC runs of each orbit are performed using the errors in distance, radial velocity, and proper motion to estimate the uncertainties in each orbital parameter. Typical uncertainties in the individual velocity components are $<1$ \vel, as shown in Fig. \ref{vsample}. It is important to stress that this is more than an order of magnitude improvement in the velocity determination compared to previous studies analysing the velocity distributions of the solar neighbourhood (e.g., \citealt{minchev2014,guiglion2015,hayden2018}). This is due to a combination of precise radial velocities from GALAH DR2, which have an accuracy of $0.1-0.2$ \vel \citep{Zwitter2018}, and the dramatic improvement in proper motions and parallaxes from \gai DR2. Velocity dispersions are measured using the equations outlined below for each \af and [Fe/H] bin, which take into account the errors in individual velocity measurements when computing the dispersion:

\begin{equation}\nonumber
\small{L(\mu,\sigma_i) = \prod_{i=1}^{N}\frac{1}{\sqrt{2\pi(\sigma_{i}^{2}+ev_{i}^{2})}}\exp\left({-\frac{1}{2}\frac{(v_i-\mu)^2}{2(\sigma_i^2+ev_i^2)}}\right)}
\end{equation}
where $\sigma_i$ is the velocity dispersion, $\mu$ is the average velocity, $v_i$ is the velocity of a given star and $ev_i$ is the error in the velocity. We minimise the log-likelihood $\Lambda\equiv2\ln{L}$ by solving where the partial derivatives of $\Lambda$ are zero, i.e. $\frac{\partial\Lambda}{\partial\mu}$ and $\frac{\partial\Lambda}{\partial\sigma_i}$, which gives:
\begin{equation}\nonumber
  \small{\sum_{i=1}^{N} \frac{v_i}{(\sigma_i^2+ev_i^2)}-\mu\sum_{i=1}^{N}\frac{1}{\sigma_i^2+ev_i^2}=0}
\end{equation}
\begin{equation}\nonumber
  \small{\sum_{i=1}^{N} \frac{(\sigma_i^2+ev_i^2)^2-(v_i-\mu)^2}{(\sigma_i^2+ev_i^2)^2}=0}
\end{equation}

see also \citet{Godwin1987,Pryor1993}. As the velocity dispersion is sensitive to extreme outliers, we remove stars more than three standard deviations away from the median in any given bin. Errors in the velocity dispersion are estimated using 10 000 bootstrap runs of the sample in each given bin. 

\section{Results}

\begin{figure*}
\centering
\includegraphics[width=6.8in]{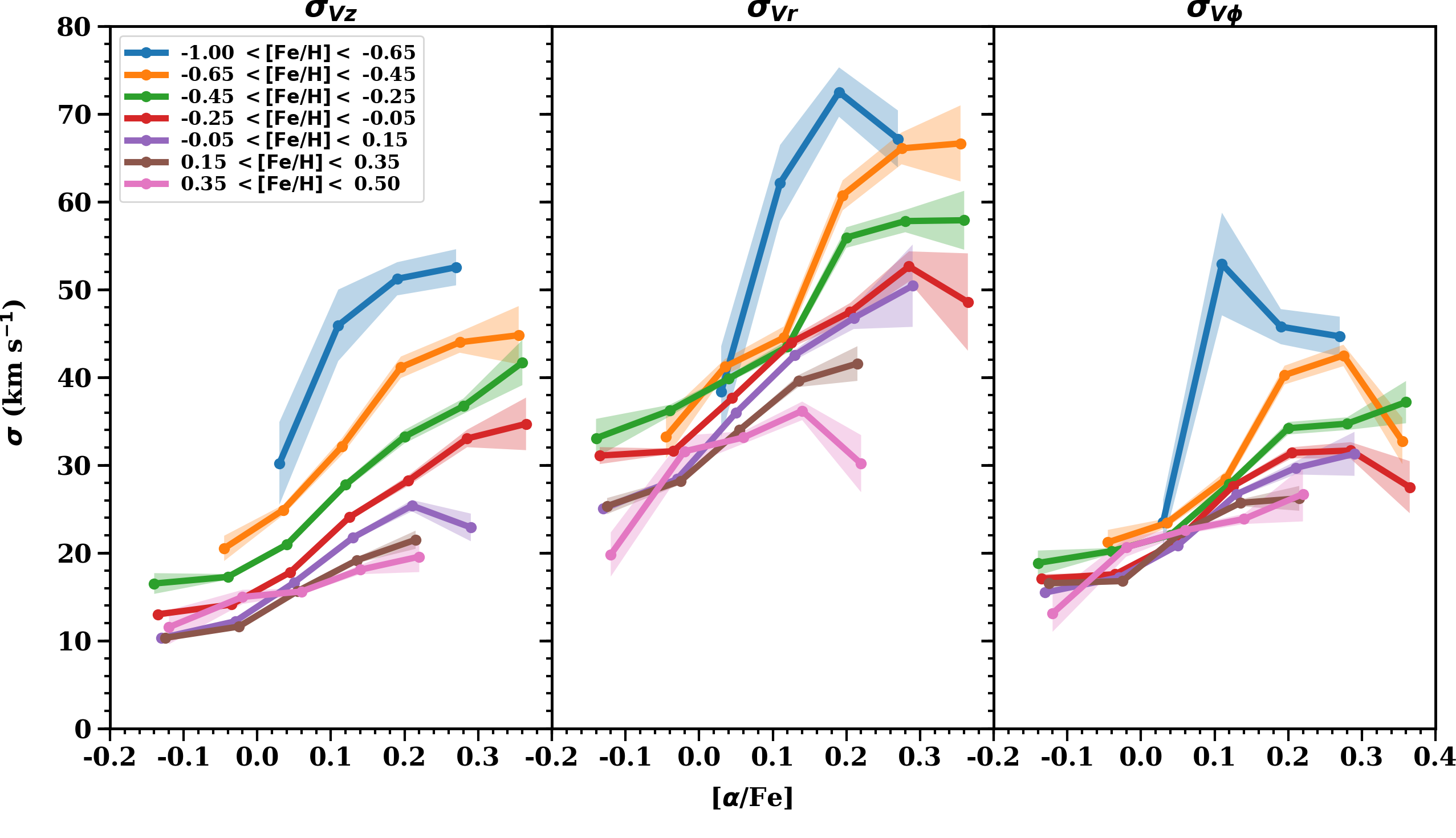}
\includegraphics[width=6.8in]{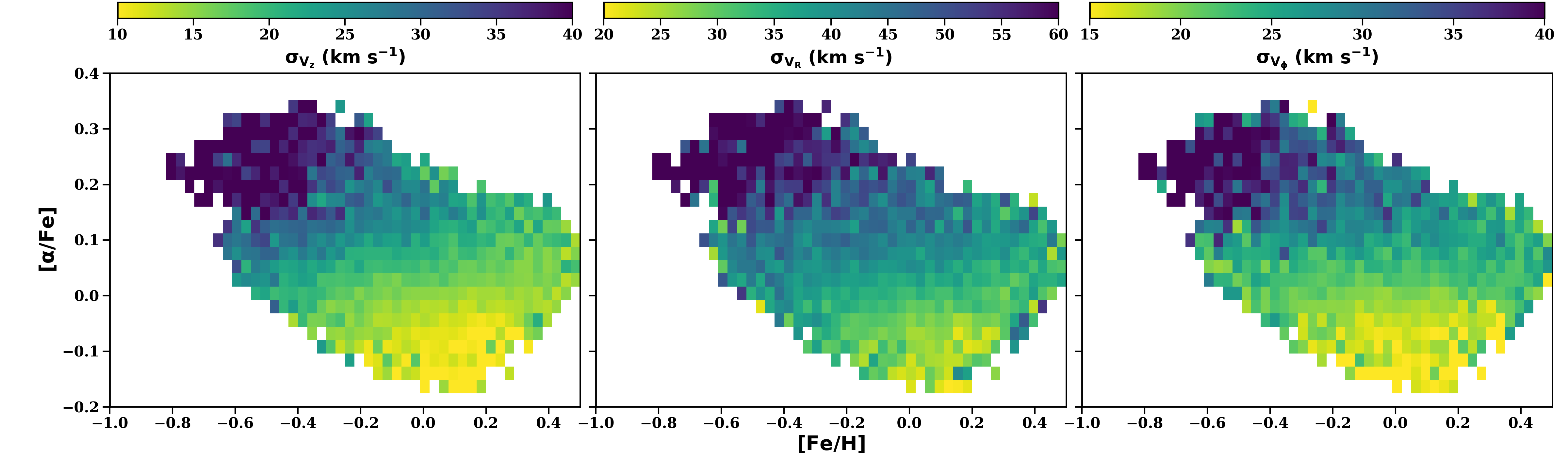}
\caption{\textbf{Top:} The velocity dispersion as a function of \af and [Fe/H] for the GALAH solar neighbourhood sample. One-$\sigma$ errors are shown by the shaded lines. There is a minimum of 20 stars per bin. \textbf{Bottom:} Similar to above, but visualised in the [$\alpha$/Fe] vs. [Fe/H] plane. There is a minimum of 10 stars per bin. The velocity dispersion varies smoothly with \af and [Fe/H], increasing with increasing \af or decreasing [Fe/H].}
\label{vdsp}
\end{figure*}

The velocity dispersion is measured in bins of \af and [Fe/H] (see Fig. \ref{afsample}) for all three components, as shown in Fig. \ref{vdsp}. There is a minimum of 20 stars per bin. We find that the vertical velocity dispersion (left panel) increases smoothly with both [Fe/H] and \afe, in fact only for the most metal-rich stellar populations does one metallicity bin intersect another for the same \afe! For stellar populations with the same [Fe/H], the velocity dispersion increases as \af increases. For populations with the same \afe, the velocity dispersion increases as [Fe/H] decreases. The overall range of the vertical velocity dispersion varies from 10 \vel for the more metal-rich and sub-solar \af abundances, to $\sim50$ \vel for the most metal-poor \afe-enhanced populations. The radial (middle panel) and tangential (right panel) velocity components follow the same trend, although with more intersection between stellar populations. The velocity dispersion is largest in the radial direction, increasing from $\sim25$ \vel to $\sim70$ \vel as \af increases and [Fe/H] decreases. The tangential velocity dispersion is generally smallest, with $\mathrm{V}_{\phi}<20$ \vel for all but the most metal-poor stars with \af$<0.1$ dex. In all components, we see a gradual flattening of the increase in velocity dispersion with increasing \af for the highest-\af stellar populations. However, we find no evidence of dramatic decrease in velocity dispersion for these high-\af stellar populations. 

\begin{figure*}
\centering
\includegraphics[width=6.0in]{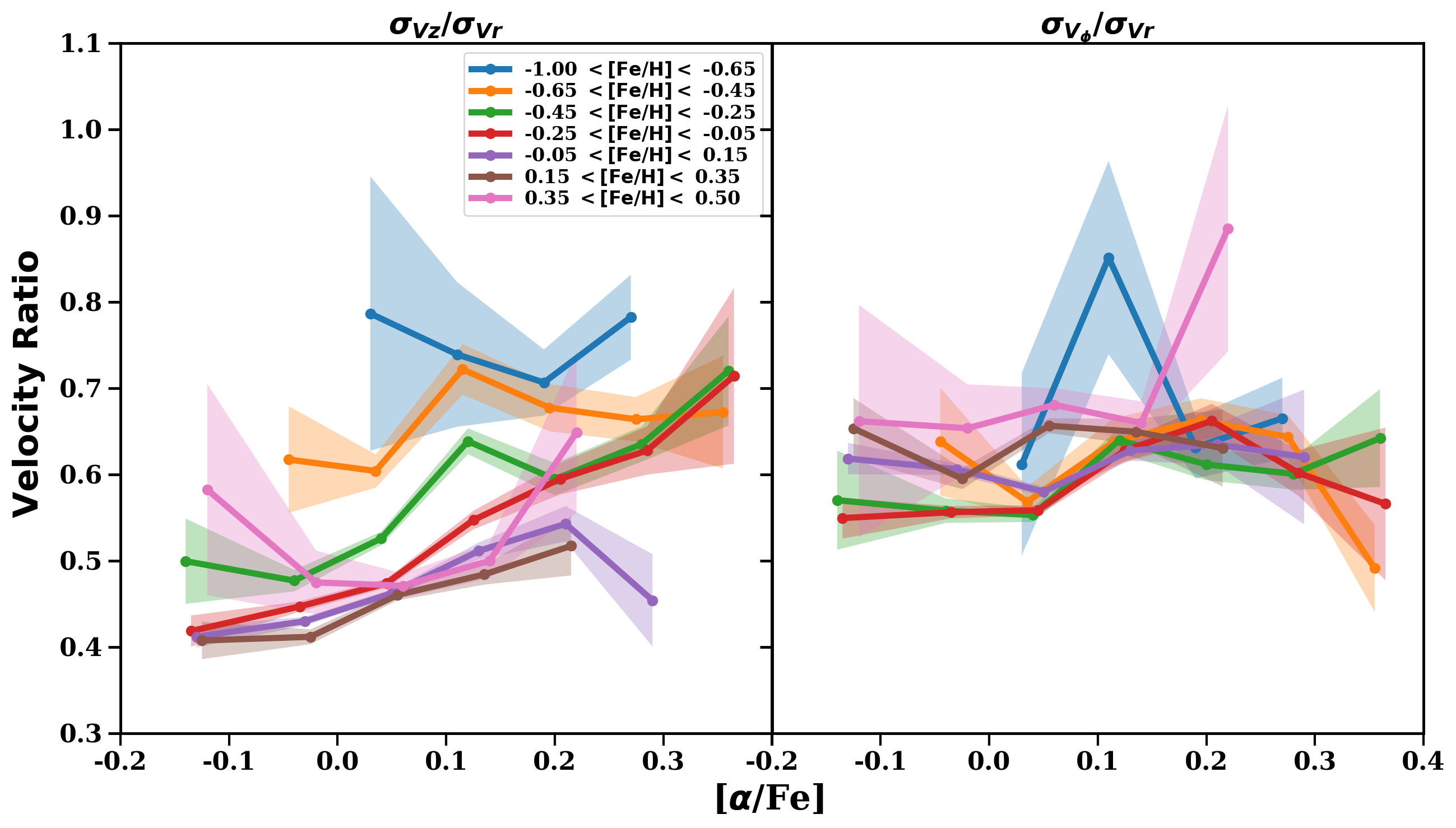}
\includegraphics[width=6.0in]{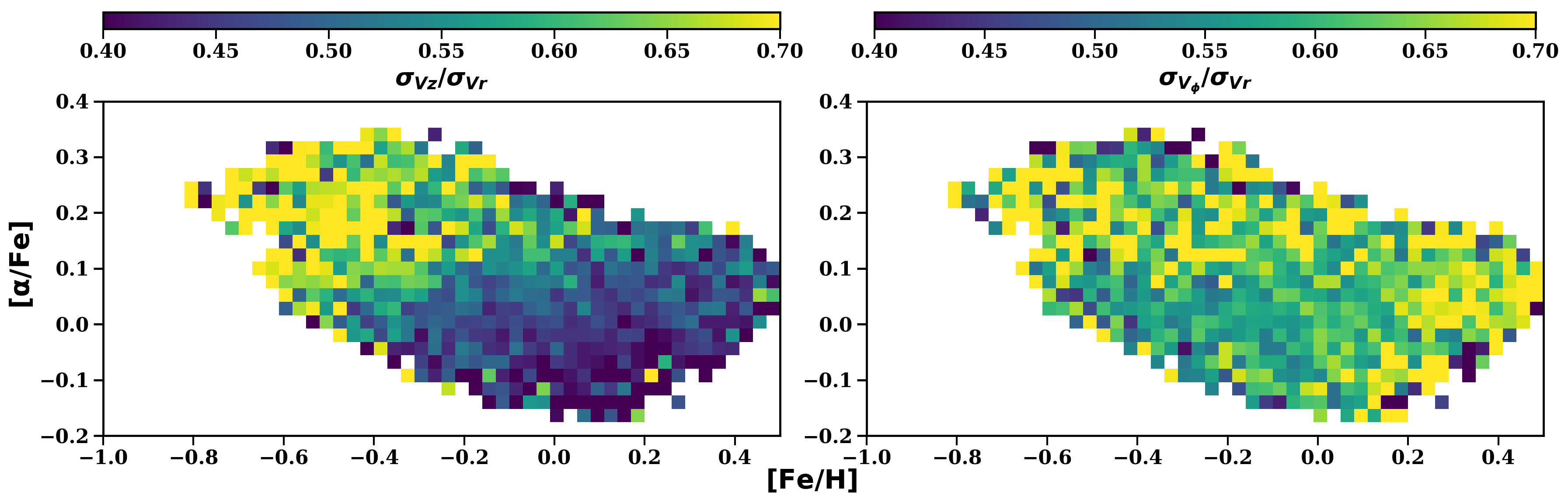}
\caption{\textbf{Top:} The ratio of the vertical velocity and rotational velocity dispersions compared to the radial velocity dispersion as a function of \af and color coded by [Fe/H]. There is a minimum of 20 stars per bin. \textbf{Bottom:} Similar to above, but in the \af vs. [Fe/H] plane, color coded by the velocity dispersion ratio. There is a minimum of 10 stars per bin. Note that the ratio of rotational velocity dispersion to radial velocity dispersion is roughly constant, so there is little structure in the bottom right panel. }
\label{vdsprat}
\end{figure*}

The ratio of the vertical and rotational velocity dispersions to the radial velocity dispersion can give insight into the various heating mechanisms that affect the different stellar populations of the disk. The ratio of $\sigma_{v_z}/\sigma_{v_R}$ and $\sigma_{v_\phi}/\sigma_{v_R}$ is shown in Fig. \ref{vdsprat} as a function of [Fe/H] and \afe. $\sigma_{v_z}/\sigma_{v_R}$ strongly varies for different stellar populations. The lowest-\af populations (chemical thin disk) have a ratio of $\sim0.4$, which gradually increases as \af increases, to $\sim0.7$ for the higher-\af populations (chemical thick disk), and even reaches 0.8 for the most metal-poor stars. There is a slight metallicity dependence, with the more metal-poor populations in general having a higher ratio of $\sigma_{v_z}/\sigma_{v_R}$, with the exception of the most metal-rich bin ([Fe/H]$>0.35$). However, this variation is less smooth than the trends identified in the individual velocity components themselves and certainly of much lower importance than the trend of increasing $\sigma_{v_z}/\sigma_{v_R}$ with \afe. In contrast, the $\sigma_{v_\phi}/\sigma_{v_R}$ ratio shows no trend with metallicity or \afe, and is roughly constant within the errors for all stellar populations at $\sim0.6$.


\begin{figure}
\centering
\includegraphics[width=3.35in]{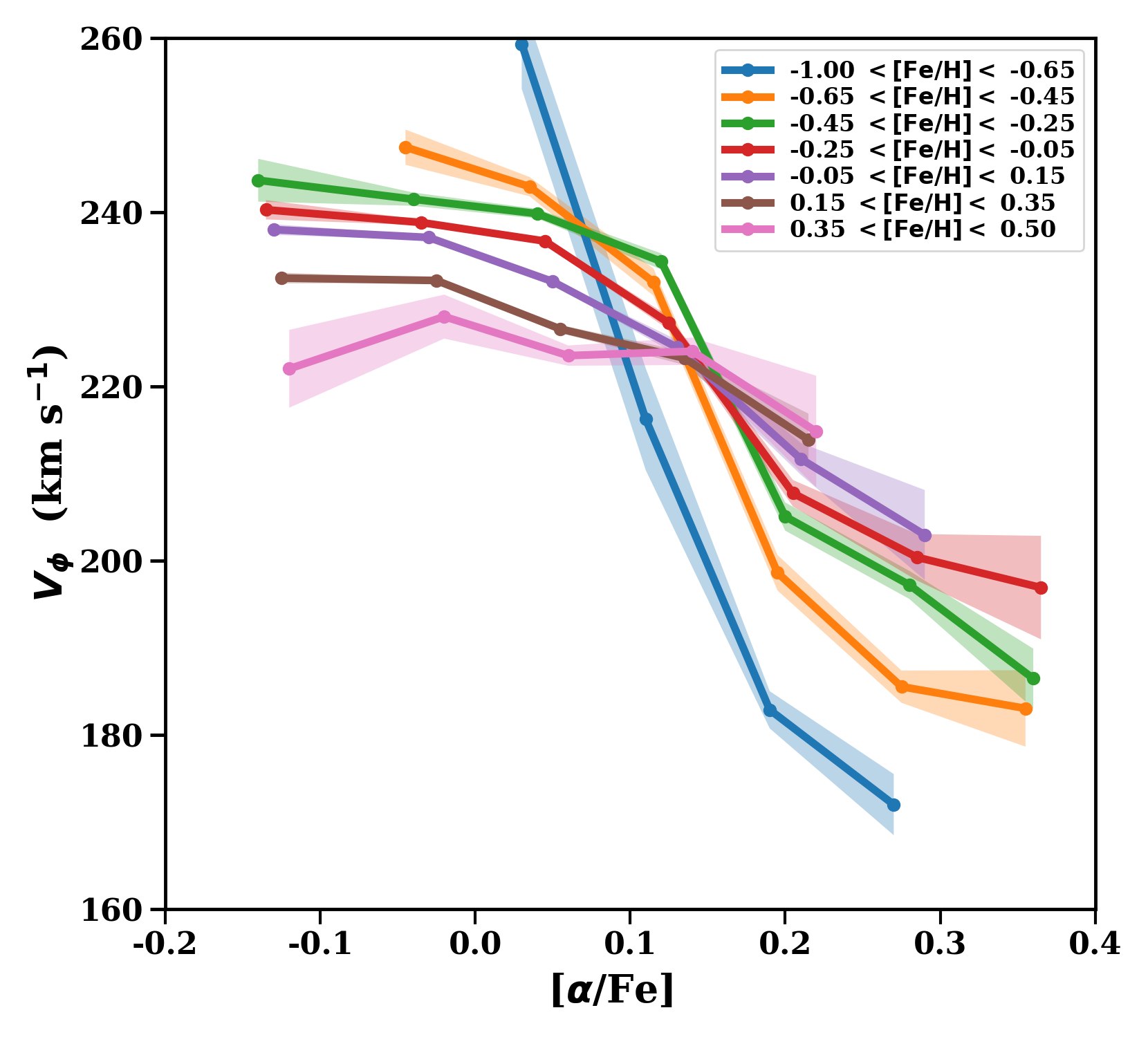}
\includegraphics[width=3.35in]{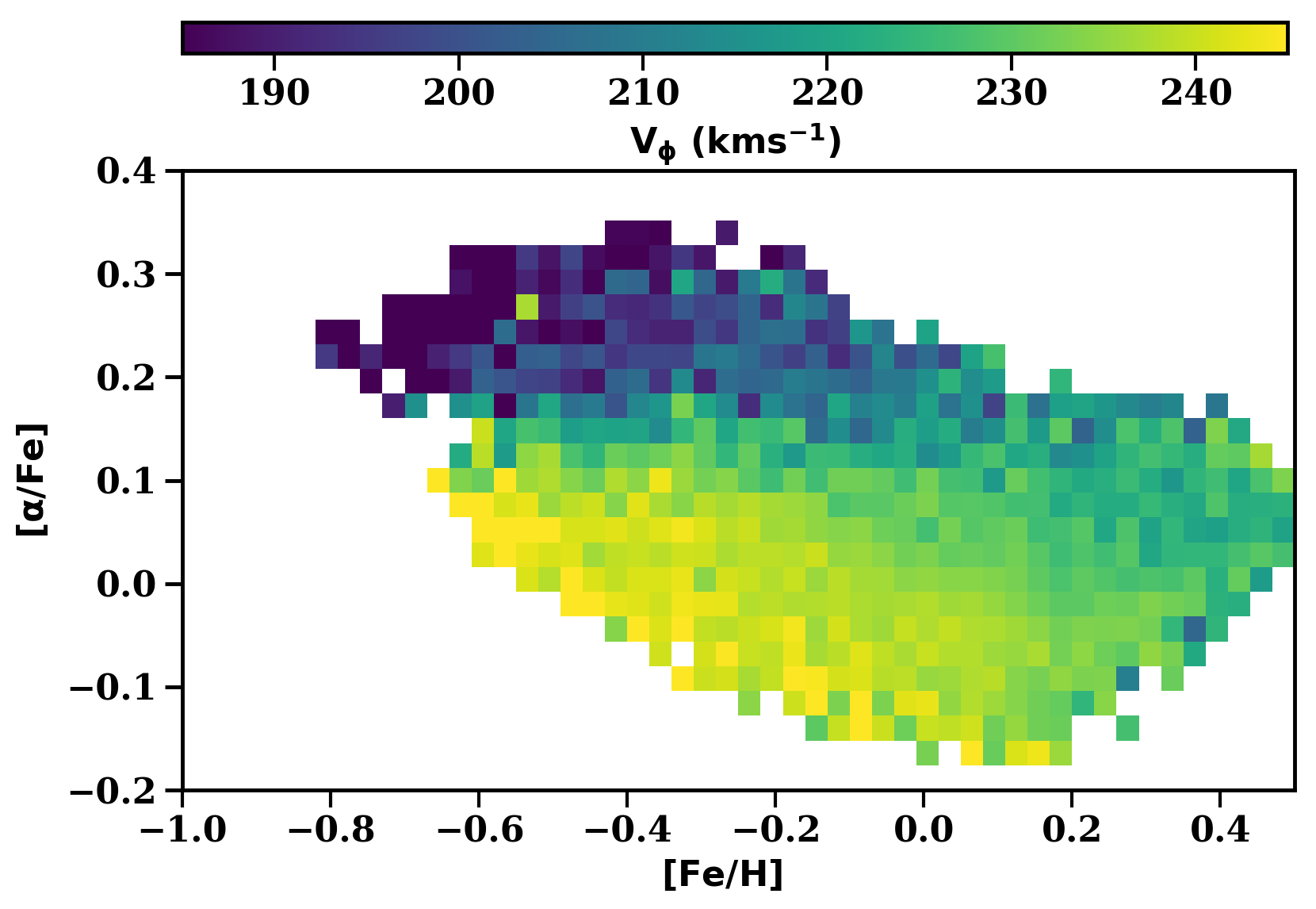}
\caption{\textbf{Top:} The median rotational velocity a function of \af and [Fe/H] for the GALAH solar neighbourhood sample. There is a minimum of 20 stars per bin. \textbf{Bottom:} The median rotational velocity in the \af vs. [Fe/H] plane. There is a minimum of 10 stars per bin.}
\label{vphi}
\end{figure}

The distribution of rotational velocities as a function of chemistry shows several interesting trends, as shown in Fig. \ref{vphi}. For stars below \af of $\sim0.1-0.2$ dex, metal-poor stars have larger rotational velocities than metal-rich stars with the same \af abundance. However, the difference in rotational velocity between the difference stellar populations is not large, $~25$ \vel for stars between $-0.65<\mathrm{[Fe/H]}<0.5$. However, these trends are reversed for higher-\af populations. For \af $\gtrsim$ $0.2$, the more metal-rich stars have higher rotational velocities than the metal-poor stars with the same \afe. The rotational velocities for these stars are also much lower at higher-\af than for lower-\af stars at the same metallicity. The velocity difference between low-\af and high-\af populations is much greater than the velocity difference of different metallicity populations at the same [$\alpha$/Fe]. 

\begin{figure}
\centering
\includegraphics[width=3.35in]{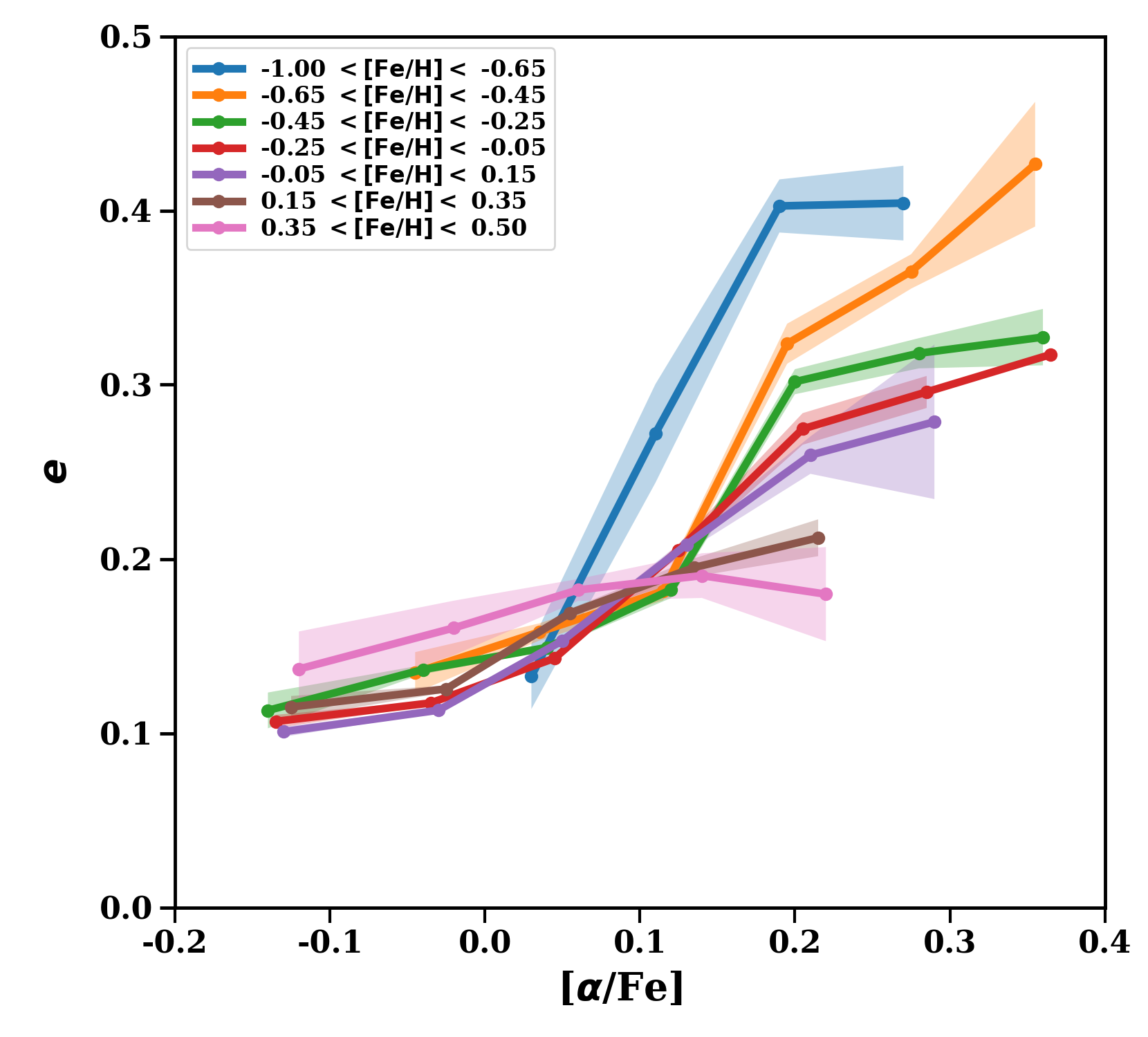}
\includegraphics[width=3.35in]{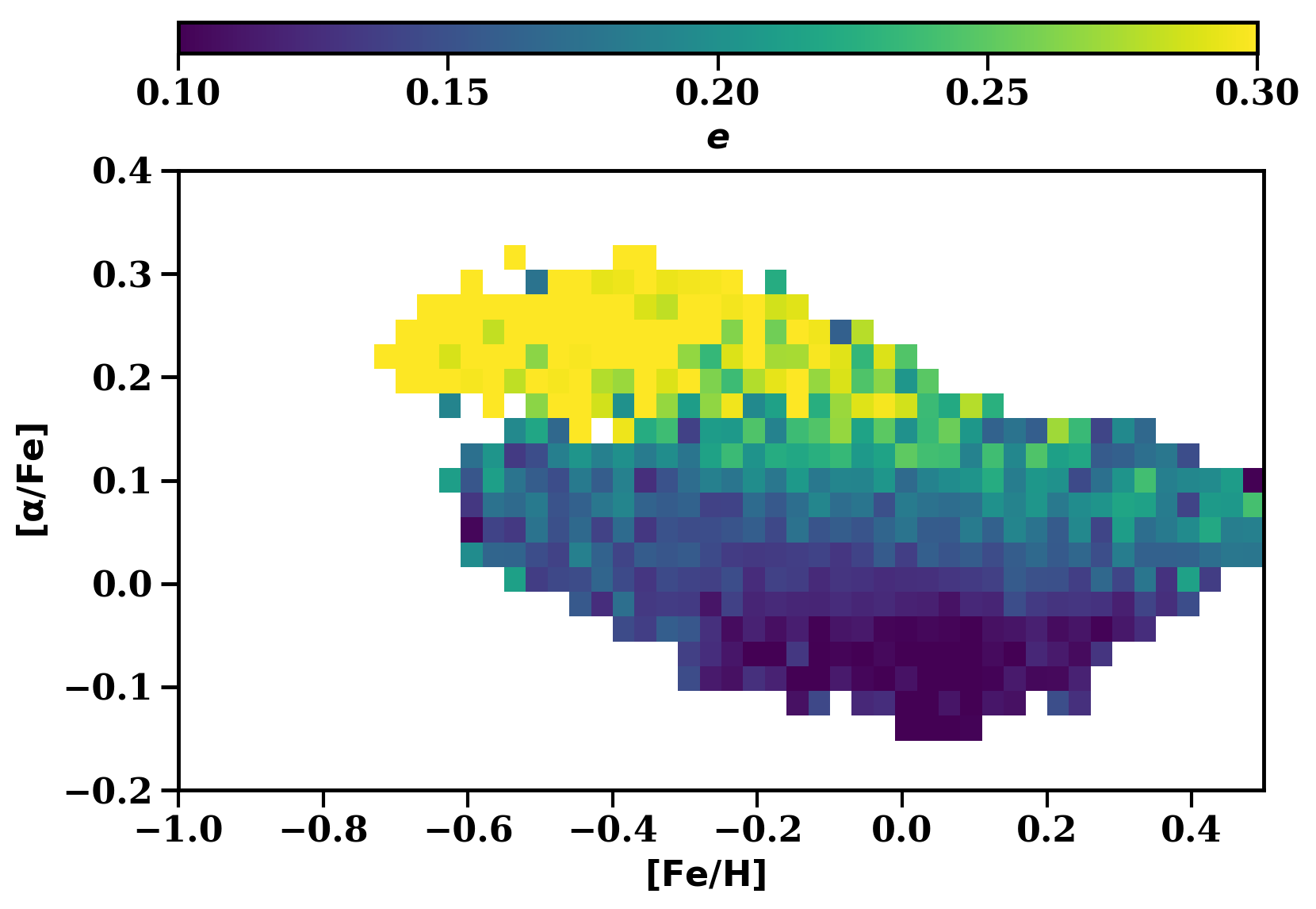}
\caption{\textbf{Top:} The median eccentricity as a function of \af and [Fe/H], with a minimum of 20 stars per bin. \textbf{Bottom:} Similar to above, but in the \af vs. [Fe/H] plane. There is a minimum of 10 stars per bin. Eccentricity increases with alpha, and for the high-\af populations metal-rich stars have lower eccentricities than metal-poor stars with the same \af on average.}
\label{ecc}
\end{figure}

The origin of the stellar populations observed in the solar neighbourhood is a particularly interesting question. The orbital properties of stars can be used to try to disentangle the dynamical mechanisms responsible for causing stars of a range of birth radii to be observed in the solar neighbourhood. The eccentricity distribution as a function of \af and [Fe/H] as shown in Fig. \ref{ecc}. For low-\af populations (\af$<0.1$ dex), the median eccentricity is relatively invariant with metallicity, with perhaps a slight increase of median eccentricity of 0.1 at the lowest-\af abundances to $\sim0.15$ at intermediate \afe. However, after this point the median eccentricities begin to increase with \af for all populations, but the increase is steepest for the more metal-poor stars, up to a maximum of $\sim0.35$ for the more metal-poor high-\af populations. These stars are on fairly elliptical orbits with large variation in radius due to epicyclic motion, and spend most of their time outside of the solar neighbourhood.

\begin{figure*}
\centering
\includegraphics[height=3.0in]{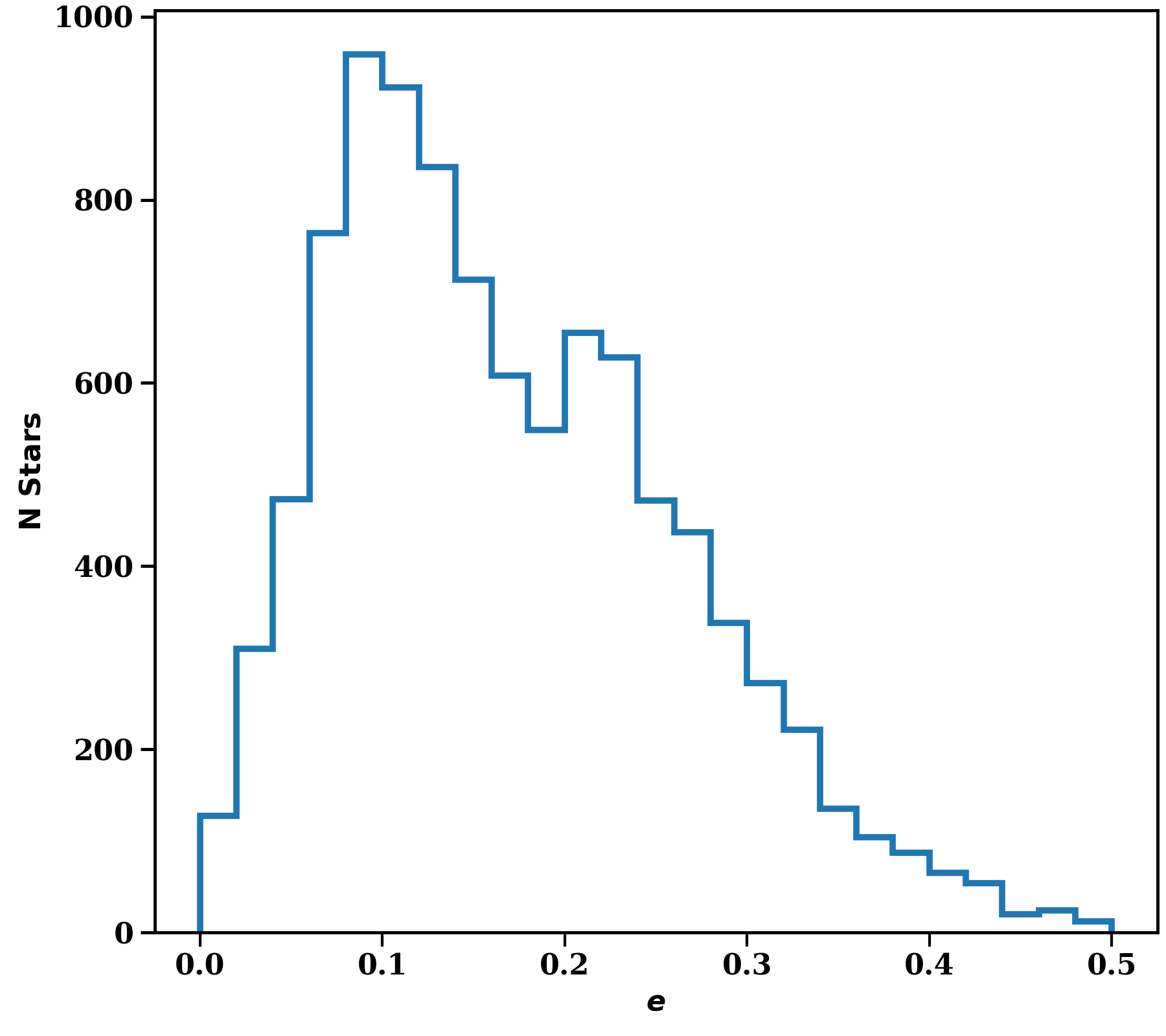}
\includegraphics[height=3.0in]{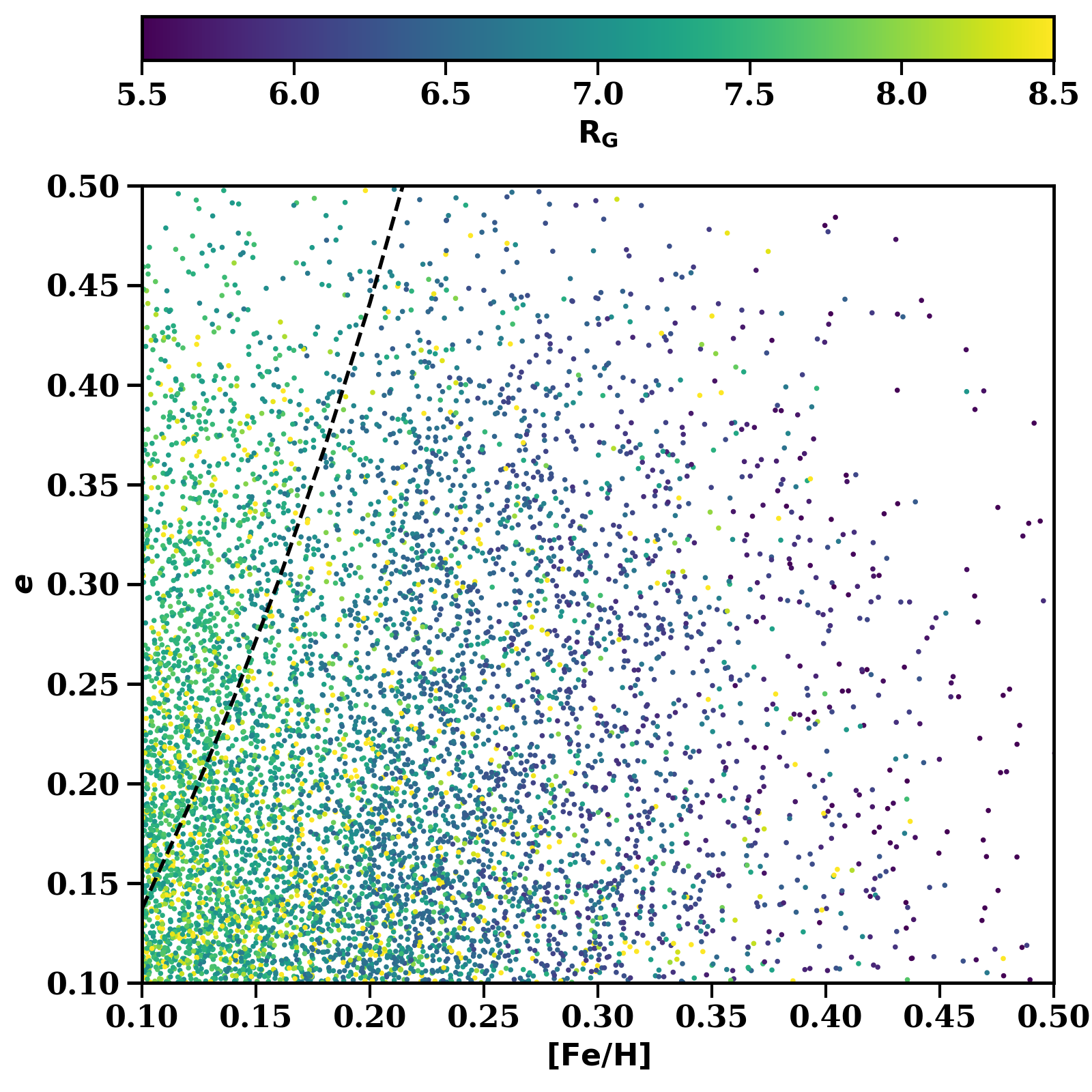}
\caption{\textbf{Left:} The distribution of orbital eccentricities for metal-rich ([Fe/H]>0.1 dex) stellar populations. \textbf{Right:} The eccentricity as a function of [Fe/H] for metal-rich stars, color-coded by the guiding radius of the orbit R$_\mathrm{G}$. The dashed line denotes the minimum required eccentricity for a star to be observed in the solar neighbourhood given it's metallicity and a radial gradient of -0.07 dex kpc$^{-1}$. Stars lying to the right/below the line require migration in order to be observed in the solar neighbourhood. }
\label{eccmetalrich}
\end{figure*}

The origin of stars with significantly higher metallicity than the local ISM in the solar neighbourhood is of particular interest. Because of their metallicities, these stars are likely to have a Galactic origin from the inner disk. As shown in the right panel of Fig. \ref{eccmetalrich}, these stars have what appears to be a bimodal distribution in eccentricity, with the dominant peak at low eccentricity ($\sim0.1$) and a second population with eccentricities centred about $e\sim0.2$. More than 70\% of these stars have eccentricities less than 0.2.  More than half (55\%) are on very circular orbits with $e<0.15$, and have relatively small epicyclic motions. For stars with [Fe/H]$>0.3$, the ratio drops slightly to 62\% with $e<0.2$ and 46\% with $e<0.15$. The overall fraction has decreased, but still a majority are on relatively circular orbits. 

The required eccentricity for a star of a given metallicity to reach the solar neighbourhood via blurring can also be estimated in a Keplerian potential using several crude assumptions. Assuming a metallicity gradient of -0.07 dex kpc$^{-1}$ (e.g.,\citealt{ luck2011a,anders2014,anders2017,duong2018}) we can get a rough estimate of the birth radius of a star given its observed metallicity and assuming the local ISM (R=8.2 kpc) is currently solar. This, combined with an assumption that the apogalacticon of the star's orbit is its present day location in the solar neighbourhood, allows us to estimate the eccentricity required to reach the solar neighbourhood as shown below:

\begin{equation}\nonumber
    e(\mathrm{Fe/H}) = \frac{R_a}{R_{g}([Fe/H])}-1
\end{equation}

Here $R_a$ is the apogalacticon of the orbit and assumed to be its present day R, and $R_g$ is the guiding radius, assumed to be its birth radius derived from it's metallicity. As all of the stars in our sample are within 500 pc of the Sun and given our previous assumptions, the $R_a$ of stars in our sample lie between 7.7 kpc and 8.7 kpc, while the guiding radius can be as small as one kpc for a star of [Fe/H]$=0.5$. This relation is shown as the dashed line in the right panel of Fig. \ref{eccmetalrich}, along with the eccentricity distribution as a function of [Fe/H] for our observed metal-rich sample. Stars lying to the left/above the line are able to reach the solar neighbourhood through blurring, while the presence of stars to the right/below the line cannot be explained by blurring alone given their measured [Fe/H] and $e$. More than 80\% of the metal-rich sample lie below the line. This finding has important implications for the roles migration and blurring play on Galactic evolution.


\section{Discussion}
The smooth variation of velocity dispersion as a function of chemistry is a surprising result, given the incredible amount of substructure in velocity space found locally (e.g., \citealt{antoja2018,quillen2018,trick2018,hunt2018,bland-hawthorn2018}). In particular, for the vertical velocity dispersion, no metallicity bin (aside from the most metal-rich stellar populations) crosses another metallicity bin in velocity dispersion as a function of \afe.  This likely implies that the stars belonging to each particular substructure in velocity space has a similar Galactic origin, as the velocity dispersion would likely show larger variations with chemistry if the various velocity substructures had a variety of different Galactic origins. The smooth variation of velocity dispersion with chemistry is similar to that observed using SEGUE observations \citep{bovy2012c}, with the velocity dispersion and scale-heights for different mono-abundance populations varying gradually with chemistry, or to that found with HARPS data which show the velocity dispersion smoothly varying with age \citep{haywood2013,hayden2017}. 

There has been evidence from previous spectroscopic studies that the signature of a major merger is visible in the velocity dispersion of local disk stars. \citet{minchev2014} found this as a sharp decrease in the velocity dispersion for the highest-\af populations using RAVE data. This signal was also observed in simulations from \citet{minchev2013}, which found that the dip in velocity dispersion for the highest-\af components corresponds to enhanced migration driven by the last major merger present in the simulation. This result was less obvious in higher-resolution spectroscopic studies \citep{guiglion2015,hayden2018}, although potentially still present within the uncertainties. However, with the advent of \gai and high-precision chemistry and radial velocity measurements from GALAH, the uncertainties in both chemistry and kinematics are significantly reduced, with the velocity precision in particular an order of magnitude better compared to previous studies. With this higher precision data, we find no evidence of a significant decrease in velocity dispersion for the highest-\af stellar populations. There appears to be a flattening in slope of the change in velocity dispersion with \af at high-\af, but not the dramatic decrease as found in the RAVE observations. The signal found in RAVE is likely the result of large uncertainties in \af, [Fe/H], and in particular the available proper motions. This does not rule out that major mergers have had an impact on the evolutionary history of the Milky Way, merely that the signal observed in previous spectroscopic studies is unable to be reproduced with higher quality observations.

The ratio of the vertical velocity dispersion to the radial velocity dispersion provides insight into different disk heating processes. Arguments of vertical heating via giant molecular cloud scattering predict a ratio of $\sigma_{v_z}/\sigma_{v_R}=0.62$ \citep{ida1993,sellwood2008,sellwood2014}. This is roughly the average for the more metal-poor and higher-\af populations observed with GALAH. However, there is clearly a trend of increasing $\sigma_{v_z}/\sigma_{v_R}$ as \af increases. \citet{aumer2009} find that the $\sigma_{v_z}/\sigma_{v_R}$ is related to the age of the stellar population, as the rate of change of the vertical velocity dispersion as a function of time is greater than that of the radial velocity dispersion. This results in an increase in the $\sigma_{v_z}/\sigma_{v_R}$ ratio for older stellar populations. To the extent that \af can be a rough proxy for age (see \citealt{haywood2013,hayden2017}), with older stellar populations having larger \af abundances, we find the same trends in the GALAH data.  \citet{sharma2014} determine the $\sigma_{v_z}/\sigma_{v_R}$ ratio for different stellar populations observed with RAVE using two different models for the velocity profile of the disk (a Gaussian distribution function and a Shu distribution function \citealt{shu1969}). For the thick disk, they find $\sigma_{v_z}/\sigma_{v_R}=0.8$ using a Shu model and $\sigma_{v_z}/\sigma_{v_R}=0.68$ for a Gaussian model, which bracket our values for the highest-\af stellar populations (classical thick disk stars) of $\sigma_{v_z}/\sigma_{v_R}$=0.7. A direct comparison to their thin disk is somewhat difficult, as their model provides an estimate of $\sigma_z/\sigma_{R}$ as a function of time. For young-intermediate age thin disk populations ($\sim3$ Gyr), they find $\sigma_{v_z}/\sigma_{v_R}= 0.44$, gradually increasing to 0.65 for the oldest thin disk stars ($\sim10$ Gyr). Our estimations for the chemical thin disk (\af$<0.15$ dex) fall inside this range, and perhaps hint that our lower-\af sample is made up of fairly young stars in general ($\sim3-5$ Gyr). 

With the advent of \gai astrometry, more recent studies have been able to place additional constraints on the local velocity dispersion. \citet{anguiano2018} measured the velocity dispersion of the local disk using a sample of high-resolution spectra and \gai DR1 astrometry and found that $\sigma_{v_\phi}/\sigma_{v_R}$ is 0.67 and 0.7 for the thin and thick disks, respectively. This is in agreement with our estimates. They also find that the the $\sigma_{v_z}/\sigma_{v_R}$ is 0.64 and 0.66, for the thin and thick disks respectively, which while not dramatically different from our estimates, does not show the trend of increasing vertical velocity dispersion relative to radial velocity dispersion with \af as observed with GALAH. Using APOGEE DR14 and \gai DR2, \citet{mackereth2019} found that the ratio of the vertical velocity dispersion to radial velocity dispersion showed large differences for thin and thick disk populations. They find that younger (lower-\afe) populations have a large spread in this ratio, driven by guiding radius and age. For their sample in the solar neighbourhood, they find that the younger populations $\sigma_{v_z}/\sigma_{v_R}$ varies between 0.4 and 0.6 as age increases, which agrees well with our measurements for lower-\af populations which show similar variation between 0.4 and 0.6 as \af increases between -0.1 and 0.1 dex. For older (higher-\afe) populations they find this ratio is roughly constant between 0.6 and 0.7. We find similar values for our high-\af populations, but a constant velocity ratio for high-\af populations is less obvious. In the APOGEE data, there is a dichotomy between young and old stellar populations, with the low-\af populations showing clear evolution in this ratio while the older populations are constant \citep{mackereth2019}. Our observations show a trend of increasing $\sigma_{v_z}/\sigma_{v_R}$ with \afe, but it is less obvious if there is a flattening of the trend for high-\af populations as is observed with the older stellar populations in APOGEE. Our results are consistent with a slight increase or a flat trend within the errors, which are large for high-\af populations.

The distribution of rotational velocities as a function of chemistry agree well with those of previous studies of the chemically selected thin and thick disk (e.g., \citealt{spagna2010,lee2011,recio-blanco2014,kordopatis2017}). These authors find that the rotational velocity for the chemical thick disk increases as metallicity increases. Conversely, for the chemical thin disk they find that rotational velocity decreases as metallicity increases. As the radial velocity dispersion is quite large for high-\af stars, these populations experience significant radial epicyclic motion which results in lower rotational velocities as these stars move outwards. As the radial velocity dispersion is smaller for the more metal-rich high-\af populations, this translates to smaller radial epicyclic motion and larger rotational velocities. These trends reverse for stellar populations with solar-\af abundances, in which the more metal-poor stars have larger radial velocity dispersions than more metal-rich stars, and higher rotational velocities. This is likely a reflection that these more metal-poor stars have guiding radii in the outer Galaxy and, for the same conservation of angular momentum arguments above, rotate faster as they move inward. For lower-\af populations, stars originating from the outer disk have generally have higher rotational velocities and lower metallicities than local populations, while stars originating in the inner disk generally have higher metallicities and lower rotational velocities than local populations. The conservation of angular momentum combined with the radial gradient of the disk nicely explains the smooth trends of rotational velocity with chemistry for populations that are \af poor. However, the difference in velocity between the solar-\af stars is relatively minor, especially compared to the higher-\af stellar populations which have significantly lower rotational velocities in general. These high-\af populations have low rotational velocities and high eccentricities, which results in large epicyclic motions and points to blurring as the likely mechanism of radial mixing that caused these stars to be observed in the solar neighbourhood. However, blurring alone may not be able to explain the presence of \af-poor populations in the solar neighbourhood. 

The fraction of stars which reach the solar neighbourhood via blurring or migration is a topic of much debate and the relative importance of these mixing processes is critical to understanding the evolutionary history of the Galaxy. The orbital properties of stars can be used to disentangle which processes dominate in the solar neighbourhood, particularly the eccentricity distribution of the most metal-rich stars observed in the solar neighbourhood. These stars must have formed in the inner Galaxy because of their metallicity, but do not have a dramatically different rotational velocity than more metal-poor populations at solar-\af abundances, implying that their guiding radii are relatively close to the solar neighbourhood. This is also reflected in the eccentricity distribution of the solar-\af populations. The majority of the most metal-rich stars in the solar neighbourhood have $e<0.2$. This is similar to results from \citet{kordopatis2015,hayden2018}, who also found that many of the most metal-rich stars in the solar neighbourhood are on fairly circular orbits. For a star with $e=0.2$ and apogalacticon in the solar neighbourhood, the guiding radius is $6.67$ kpc and the deviation from this guiding radius is only $\sim1.33$ kpc. This means the eccentricity must be quite high for a very metal-rich star coming from the inner Galaxy to reach the solar neighbourhood. More than 70\% of the stars have eccentricities $e<0.2$, implying their deviation from their guiding radius is $<1$ kpc. While there are several significant assumptions required to generate the relation shown by the dashed line in Fig. \ref{eccmetalrich}, it clearly demonstrates that blurring is insufficient to explain the presence of the majority of these metal-rich stars in the solar neighbourhood. Even with the simple assumptions made here, it is almost impossible for stars with [Fe/H]$>0.3$ to reach the solar neighbourhood via blurring, as the orbit must be hyperbolic given their very high metallicities and the observed radial gradient of -0.07 dex kpc$^{-1}$. For these stars to reach the solar neighbourhood, significant angular momentum transfer (migration) must take place. Previous studies have argued that observations of the solar neighbourhood can be explained by blurring alone (e.g., \citealt{haywood2013}), but our results here are at odds with that interpretation. 

It is important to note that different radial mixing processes are not mutually exclusive. A star that migrates can subsequently be heated; while a star that has had its orbit heated can migrate, particularly if that star has small vertical motion as the efficiency of migration from non-axisymmetric structure is largest for stars in the plane \citep{solway2012,vera-ciro2016,daniel2018}. The fact that the overwhelming majority of the most metal-rich stars have eccentricities incompatible with blurring alone to reach the solar neighbourhood shows that this process is a necessary component in realistic models of the chemodynamic evolution of the Galaxy. Both blurring and churning play a role in shaping the evolution of the Milky Way.

\section{Conclusion}

We analysed the chemodynamic structure of the local solar volume (d$<500$ pc) using 62 814 stars in GALAH DR2 in conjunction with astrometric information from \gai DR2. We find that the velocity dispersion generally increases smoothly as [Fe/H] decreases or \af increases, with the vertical velocity dispersion in particular showing no large discontinuity or even crossing in velocity space of different stellar populations. This is in stark contrast to the U-V velocity and action planes, which show several over densities and significant substructure. There is no evidence of a downturn in velocity dispersion for the highest-\af stellar populations as previously observed with lower quality data, which indicates previous results were likely spurious. This result highlights the power that \gai and large-scale spectroscopic surveys such as GALAH bring in disentangling the structure of the disk, dramatically increasing the precision and accuracy of velocity and orbital determination compared to previous studies. 

The stars that are currently in the solar neighbourhood come from a large range of birth radii. How these stars reach the solar neighbourhood holds important clues on how the Milky Way disk has evolved with time. Populations with higher-\af abundance tend to have large vertical velocity dispersion and low rotational velocities, with large asymmetric drift, implying an origin in the inner Galaxy and blurring as the dominant radial mixing process to observe these stars in the solar neighbourhood. Conversely, for the solar-\af populations, lower-metallicity stars have much larger rotational velocities on average than the LSR, implying a Galactic origin in the outer galaxy. Expanding this analysis to the orbital properties for stars that have metallicities much higher than the local ISM can also shed light onto different radial mixing processes of the disk. Based on the eccentricity distribution of these metal-rich stars ([Fe/H]$>0.1$ dex), we find that for the majority blurring alone is unable to explain the presence of these stars near the solar position. The majority of these stars are on fairly circular orbits and simply never reach the Galactic radii at which they formed based on their high-metallicities. For these stars, churning/migration is required and therefore is the dominant radial mixing process to observe these populations in the solar neighbourhood. These observations highlight the importance that migration plays in the structure and evolution of the Milky Way, realistic models of the chemodynamic history of the Galaxy must include migration in order to adequately explain observations. 

\section{Acknowledgements}
This work has made use of data from the European Space
Agency (ESA) mission \gai (https://www.cosmos.esa.
int/gaia), processed by the \gai Data Processing and
Analysis Consortium (DPAC, https://www.cosmos.esa.
int/web/gaia/dpac/consortium). This work is also based
on data acquired from the Australian Astronomical Telescope.
We acknowledge the traditional owners of the land
on which the AAT stands, the Gamilaraay people, and pay
our respects to elders past and present. This research was supported by the Australian Research Council Centre of Excellence for All Sky Astrophysics in 3 Dimensions (ASTRO 3D), through project number CE170100013.

In addition to ASTRO3D, MRH received support from ARC DP grant DP160103747. TZ acknowledges financial support of the Slovenian Research Agency (research core funding No. P1-0188). 
JBH is supported by an Australian Laureate Fellowship from the ARC.
SLM acknowledges support from the Australian Research Council through grant DP180101791.




\bibliographystyle{mnras}
\bibliography{zref} 

\bsp	
\label{lastpage}
\end{document}